# Load Profile Inpainting for Missing Load Data Restoration and Baseline Estimation

Yiyan Li, *Member, IEEE*, Lidong Song, *Student Member, IEEE*, Yi Hu, *Student Member, IEEE*, Hanpyo Lee, *Student Member, IEEE*, Di Wu, *Member, IEEE*, PJ Rehm, and Ning Lu, *Fellow, IEEE*

***Abstract*—This paper introduces a Generative Adversarial Nets (GAN) based, *Load Profile Inpainting Network* (Load-PIN) for restoring missing load data segments and estimating the baseline for a demand response event. The inputs are time series load data before and after the inpainting period together with explanatory variables (e.g., weather data). We propose a Generator structure consisting of a coarse network and a fine-tuning network. The coarse network provides an initial estimation of the data segment in the inpainting period. The fine-tuning network consists of self-attention blocks and gated convolution layers for adjusting the initial estimations. Loss functions are specially designed for the fine-tuning and the discriminator networks to enhance both the point-to-point accuracy and realisticness of the results. We test the Load-PIN on three real-world data sets for two applications: patching missing data and deriving baselines of conservation voltage reduction (CVR) events. We benchmark the performance of Load-PIN with five existing deep-learning methods. Our simulation results show that, compared with the state-of-the-art methods, Load-PIN can handle varying-length missing data events and achieve 15-30% accuracy improvement.**

## I. Introduction

MISING data is a common issue in distribution system load profile processing. Often times, missing data are caused by temporarily communication lost with the equipment. Statistics show that approximately 70% of missing data segments in power distribution systems are less than 4 hours. Thus, restoring short missing data segments is critical for improving the data utilization rate and providing high-quality data sets to down-stream data processing tasks.

Moreover, algorithms used for restoring missing data can also be used to identify the operational baseline of demand response (DR) programs. For example, Conservation Voltage Reduction (CVR) is widely used by utilities for peak load reduction [1]. During a CVR event, system voltage at the substation bus will be reduced by 2-4% to achieve load reduction. To quantify the CVR-caused load reduction, it is very important for utility engineers to accurately estimate what the original load profile (i.e., the baseline) during a CVR event would have been had the bus voltage not been reduced. Because the pre- and post- CVR load profiles represent customer consumption patterns under normal system voltage, uncovering the CVR baseline is equivalent to restoring missing data in the CVR period. Because baseline estimation is essential to DR performance evaluation, inpainting the would-have-been load profile during a DR event is highly valuable to load service providers.

Existing missing data restoration methods for load profile inpainting are categorized into *model-based* and *data-driven*. Model-based methods use physical system models to simulate responses to external disturbances in hope of restoring missing data segments. For example, to estimate the CVR baseline, researchers use the distribution system topology and load models to predict load changes when system voltage changes [2]-[6]. In general, model-based methods require accurate distribution system models. However, in practice, distribution system models are incomplete and inaccurate due to topology changes and lack of measurements. Moreover, load composition varies with respect to the time-of-the-day while customer consumption patterns shift constantly due to occupancy and weather conditions. Therefore, it is often times infeasible for utilities to apply model-based methods for restoring missing data segments.

Thus, the data-driven approach is the dominant approach for missing data restoration. Data-driven methods can be further categorized into three approaches: *similarity-based, regression-based*, and *generative-based*. The similarity-based approach groups load profiles by day type, weather conditions, and shape characteristics of load profiles. The missing data segments are restored by referencing to the data on the load profiles having the best similarity match. Similarity-based methods are straightforward, easy to implement and explainable, therefore are widely used in field implementation [7]-[10]. However, in many cases, similarity metrics are normally defined by human analysts and can be based on the weighted average of many factors (e.g., weather, time, geographical conditions, and load types). This makes the accuracy of the method dependent on subjective selections of similarity metrics and weights.

Regression-based methods include linear regression [11],

This research is supported by the U.S. Department of Energy's Office of Energy Efficiency and Renewable Energy (EERE) under the Solar Energy Technologies Office Award Number DE-EE0008770.

Yiyan Li is with the College of Smart Energy, Shanghai Jiao Tong University, Shanghai, 200240, China, and also with the Electrical and Computer Engineering Department, Future Renewable Energy Delivery and Management Systems Center, North Carolina State University, Raleigh, NC 27606 USA (email: yiyan.li@sjtu.edu.cn).

Lidong Song (corresponding), Yi Hu, Han pyo Lee, and Ning Lu are with the Electrical & Computer Engineering Department, Future Renewable Energy Delivery and Management (FREEDM) Systems Center, North Carolina State University, Raleigh, NC 27606 USA. (e-mails: lsong4@ncsu.edu, yhu28@ncsu.edu, hlee39@ncsu.edu, nlu2@ncsu.edu).

PJ Rehm is with ElectriCities (e-mail: prehm@electricities.org ). Di Wu is with Pacific Northwest National Laboratory (e-mail: Di.Wu@pnnl.gov ).

Long Short Term Memory (LSTM) [12], Stacked Autoencoder (SAE) [13], Gaussian Regression [14], Support Vector Regression (SVR) [15][16], etc. Regression-based methods usually achieve higher estimation accuracy compared to the similar day approach because of their nonlinear learning capabilities, especially when using deep-learning models. However, compared with similarity-based methods, the deep-learning based methods are less explainable and having higher computing costs. In recent years, hybrid solutions combining multiple regression models [17]-[19] are proposed for baseline estimation or missing data restoration.

The main drawbacks of traditional regression-based models are that the data format of the input and output is required to be fixed. However, in practice, the duration of missing data (model output) varies from minutes to several hours, and the length and number of available measurements (model input) also vary case by case. To cope with the varying-length cases, traditional methods need either increase the output window to cover the longest event or train separate models for different scenarios. This increases model complexity with additional computing and deployment costs, and the model performance can hardly be guaranteed due to information losses.

In this paper, we propose a third approach for missing data restoration: the Generative Adversarial Nets (GAN) based approach. Studies of using GAN to solve the missing data restoration problem is still in infancy. In [20][21], the authors discuss the basic theory of GAN in restoring missing data, while in [22][23], the authors implement a GAN-based method in power domain to restore the grid measurement data and the PV profiles. Inspired by image inpainting, we develop a Load Profile Inpainting Network (Load-PIN) using GAN [24] as the basic structure of a highly accurate and flexible missing data restoration framework for recovering missing data on load profiles and estimating baselines for demand response events. The generator consists of a coarse network and a fine-tuning network. Initially, the bidirectional time series load data before and after the missing data segment together with the explanatory variables are fed into the coarse network to obtain an initial estimation for the missing part. Next, initial estimations are sent to the fine-tuning network consisting of Gated Convolution layers [25] and Multi-head self-attention blocks [26] to improve accuracy. The generator network is trained under the guidance of the discriminator with specially designed loss functions.

The main contribution of this paper is the development of the Load-PIN model for missing data restoration and baseline estimation. The most distinct feature of the Load-PIN model is its flexibility in restoring variable-length data segments and its superior accuracy compared with the state-of-the-art methods. To the best of the authors' knowledge, the generative approach has not yet been developed for patching missing smart meter data and for demand response baseline identification.

The rest of the paper are organized as follows: Section II formulates the missing data restoration and CVR baseload estimation problem, and introduce the proposed Load-PIN model, Section III demonstrates the case study results, and Section IV concludes this paper.

## II. Methodology

In this section, we first illustrate the problem formulation of load profile inpainting and the background of the GAN based approach. Then, we present the proposed Load-PIN framework including the 2-stage GAN generator structure and the loss function design.

### A. Problem Formulation of Load Profile Inpainting

Denote a historical time series matrix of load as $\mathbf{Y} = [y_1, y_2, \dots, y_L]$, where $L$ is the length of the time series. Denote the explanatory variables $\mathbf{X}$ as

$$\mathbf{X} = \begin{bmatrix} x_1^1 & x_2^1 & \cdots & x_L^1 \\ x_1^2 & x_2^2 & \cdots & x_L^2 \\ \vdots & \vdots & \ddots & \vdots \\ x_1^E & x_2^E & \cdots & x_L^E \end{bmatrix} \tag{1}$$

where $E$ is the number of explanatory variables.

Define an event to be a missing data segment (see Fig. 1(a)) or an unknown DR baseline (see Fig. 1(b)). Assume there are $N$ events in $\mathbf{Y}$ and the duration of the $i^{\text{th}}$ event (i.e. the $i^{\text{th}}$ inpainting period) is $T^i_{event}$. The inpainting duration vector, $\mathbf{T}_{event}$, is

$$\mathbf{T}_{event} = [T^1_{event}, T^2_{event}, \dots, T^N_{event}] \tag{2}$$

while the $i^{\text{th}}$ restored data segment, $\hat{\mathbf{Y}}^i_{event}$, is

$$\hat{\mathbf{Y}}^i_{event} = \left[\hat{y}^i_1, \hat{y}^i_2, \dots, \hat{y}^i_{T^i_{event}}\right] \tag{3}$$

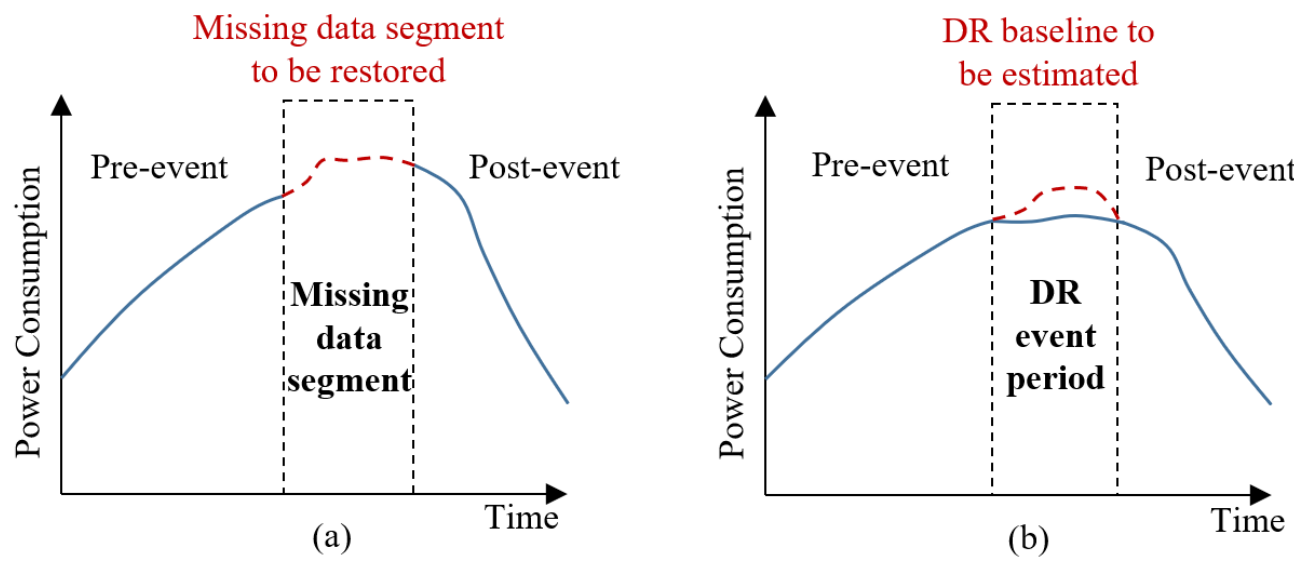

Fig. 1. Two basic event types: (a) mising data restoration, and (b) DR baseline identification. Blue solid lines are field measurements and red dotted lines are uncovered data segments.

As shown in Fig. 2, we divide the load profile containing the $i^{\text{th}}$ event into three periods: $[\mathbf{X}^i_{event}, \mathbf{Y}^i_{event}]$ as the inpainting data period, $[\mathbf{X}^i_{pre}, \mathbf{Y}^i_{pre}]$ as the pre-event period, and $[\mathbf{X}^i_{post}, \mathbf{Y}^i_{post}]$ as the post-event period.

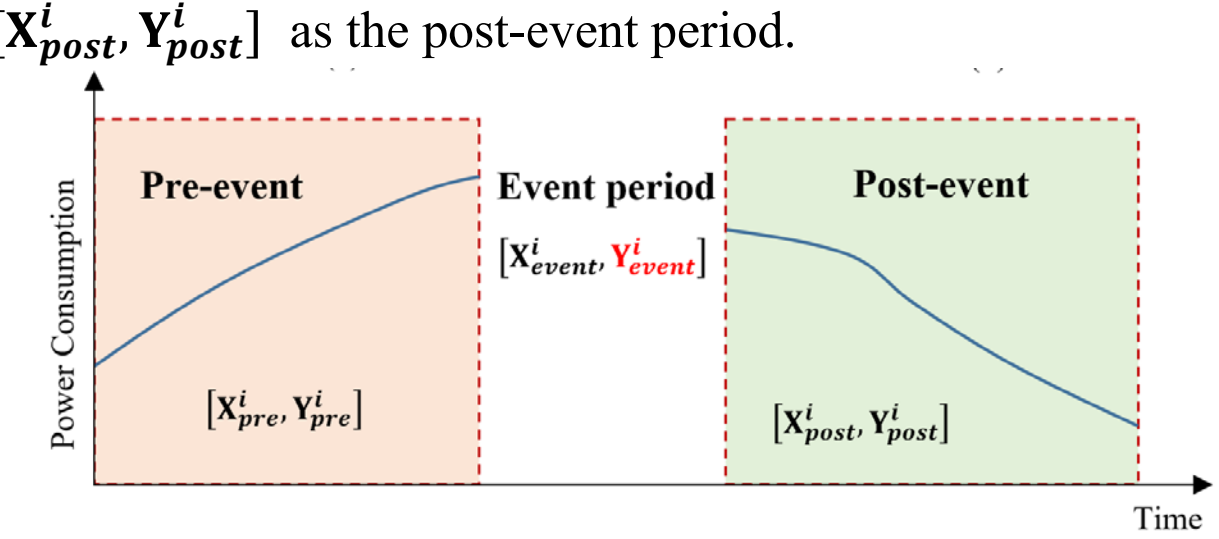

Fig. 2. An illustration of the division of the load profile containg an event.

Thus, a load profile inpainting problem can be described as

$$\hat{\mathbf{Y}}^i_{event} = f_\theta\left(\mathbf{X}^i_{pre}, \mathbf{Y}^i_{pre}, \mathbf{X}^i_{event}, \mathbf{X}^i_{post}, \mathbf{Y}^i_{post}\right) \tag{4}$$

where $f_\theta$ is the mapping function.

In the similarity-based approach, pre- and post- event data are mainly used for identifying similar days. Traditional forecasting-based methods use only pre-event data, $\mathbf{X}_{pre}^{i}$ and $\mathbf{Y}_{pre}^{i}$, as model inputs to forecast $\hat{\mathbf{Y}}_{event}^{i}$. Most correlation-based methods take $\mathbf{X}_{event}^{i}$ as inputs to build the mapping function between $\mathbf{Y}$ and $\mathbf{X}$ (e.g., temperature and voltage) for estimating $\hat{\mathbf{Y}}_{event}^{i}$. Thus, one of the main drawbacks of the state-of-the-art methods is that the information contained in all five available data sets (i.e., $[\mathbf{X}_{pre}^{i}, \mathbf{Y}_{pre}^{i}, \mathbf{X}_{event}^{i}, \mathbf{X}_{post}^{i}, \mathbf{Y}_{post}^{i}]$) have not yet been fully utilized in one shot. To resolve this deficiency, we take the GAN based approach to use all 5 available data sets as inputs to $f_\theta$ for predicting $\hat{\mathbf{Y}}_{event}^{i}$.

### B. GAN-based Approach

As shown in Fig. 3, a GAN model consists of a generator network ($G$) and a discriminator network ($D$). The input of the generator is a latent vector $\mathbf{z}$, often a Gaussian noise. The generated data, $G(\mathbf{z})$, along with the actual data, $\mathbf{x}$, are then passed to discriminator $D$. The goal of $D$ is to distinguish real data sets from the fake ones. The training of a GAN model is an iterative, adversarial process: $G$ tries to generate samples $G(\mathbf{z})$ to fool $D$; $D$ learns to identify $G(\mathbf{z})$ from $\mathbf{x}$ by assigning greater probabilities to $\mathbf{x}$ and smaller ones to $G(\mathbf{z})$. As introduced in [24], this process is formulated as a minimax game

$$\min_{G}\max_{D} R(D,G) = \mathbb{E}_{\mathbf{x}\sim p_{(\mathbf{x})}}[\log D(\mathbf{x})] + \mathbb{E}_{\mathbf{z}\sim p_{(\mathbf{z})}}[\log(1-D(G(\mathbf{z})))] \quad (5)$$

where $R(D,G)$ is the reward function, $p(\mathbf{x})$ and $p(\mathbf{z})$ are the probability distributions of training data and latent vector, $\mathbb{E}$ is the expectation operator.

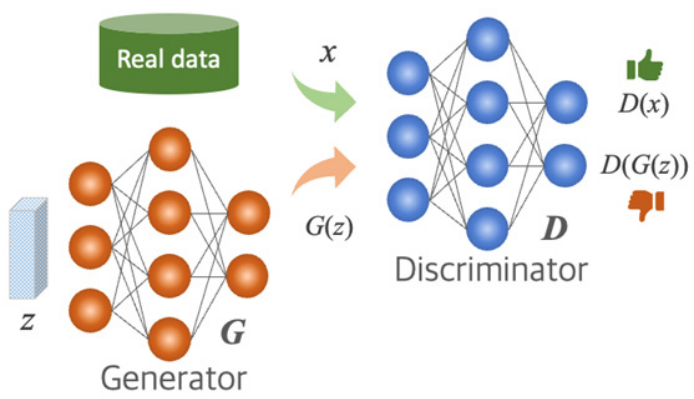

Fig. 3. The original GAN model.

### C. Work Flow of the Load-PIN Framework

The Load-PIN framework is illustrated in Fig. 4. The model input $\mathbf{z}$ has three parts: 24-hour load and temperature profiles, and a Boolean mask indicating the event period as one and the normal period as zero. The load data resolution varies from 1-minute to 15-minute and the missing data duration, $T_{event}$, is less than 4 hours. The generator contains two stages: a coarse network for initial estimation and a fine-tuning network for polishing. The discriminator is a deep convolutional network with specially designed loss functions.

#### 1) Training Sample Generation

The goal of the sample generation process is to generate samples evenly distributed in yearly load profiles so that the trained model will not be biased by factors such as the time-of-the day and season-of-the-year. To achieve this goal, we slide a 24-hour moving window over the yearly historical load profiles, as shown in Fig. 5. In this paper, the time shift ($\triangle t$) of the moving window is one hour.

There are two considerations when creating training samples. *First*, to train and test the Load-PIN model, the training and testing samples should be generated from load profiles containing no CVR events. This is because, during a CVR event, system voltages are reduced by 2-4% for load reduction, making the load profile under the normal system voltage (the ground truth) unknown to the analyst to evaluate the model performance. *Second*, the missing data segment (indicated by the mask) will be positioned at the center of the 24-hour window so that the pre- and post- event data are equal in length.

After the Load-PIN model is trained, we apply it to CVR samples for CVR baseline identification. A CVR sample is a 24-hour load profile with the CVR event being placed in the middle of the 24-hour window (See Fig. 5).

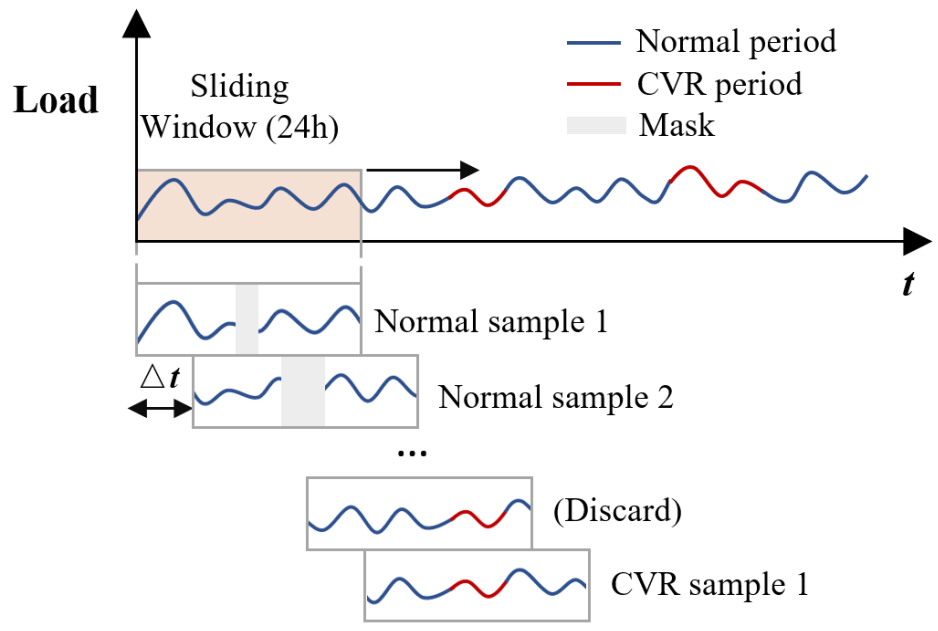

Fig. 5. Training sample generation process.

Denote $N_{train}$ , $N_{test}$, and $N_{CVR}$ as the total number of training samples, testing samples, and CVR samples, respectively. For the $i^{\text{th}}$ normal sample, a mask of variable length (but less than 4-hour) is placed in the middle of the 24-hour window. The data being masked is the ground truth ($\mathbf{Y}_{event}^{i}$) of the forecasted missing data segment ($\hat{\mathbf{Y}}_{event}^{i}$). Before and after the mask are the pre-event and post-event load segments (i.e., $\mathbf{Y}_{pre}^{i}, \mathbf{Y}_{post}^{i}$). Meanwhile, load segments are paired with their corresponding temperature profiles as the explanatory variable (i.e.,$\mathbf{X}_{pre}^{i}, \mathbf{X}_{event}^{i}, \mathbf{X}_{post}^{i}$). After $\hat{\mathbf{Y}}_{event}^{i}$ is estimated using (4), we can optimize the model parameters by

$$\theta^{*} = \arg\min_{\theta} \sum_{i=1}^{N_{train}} \left\| \mathbf{Y}_{event}^{i} - \hat{\mathbf{Y}}_{event}^{i} \right\|_2^2 \quad (6)$$

#### 2) Generator Network: 1st-Stage Coarse Network

In the first stage, we employ the Gated Convolution Network (GCN) [25] to formulate an encoder-decoder structure to restore the masked segments. Compared with conventional convolution neural network (CNN), GCN adds an additional feature-wise gating control mechanism, which learns soft masks automatically to compute the hidden layer features as

$$h_l(\mathbf{X}) = \varphi(\mathbf{X}\cdot\mathbf{W}+\mathbf{b}) \otimes \sigma(\mathbf{X}\cdot\mathbf{U}+\mathbf{c}) \quad (7)$$

where $\mathbf{X}$ is the input of layer $h_l$, $\mathbf{W}$, $\mathbf{b}$, $\mathbf{U}$, $\mathbf{c}$ are trainable parameters. $\sigma$ is the sigmoid function, $\varphi$ can be any activation function (e.g., ReLU and LeakyReLU), and $\otimes$ is the element-wise product between matrices. This equation demonstrates that different weights of filters are applied at different temporal points to produce the output, resulting in a dynamic feature selection mechanism for each channel and each time point. Besides, Gated Transpose Convolution Network (GTCN) [27] layers are introduced to recover the estimated daily profiles from the GCN output.

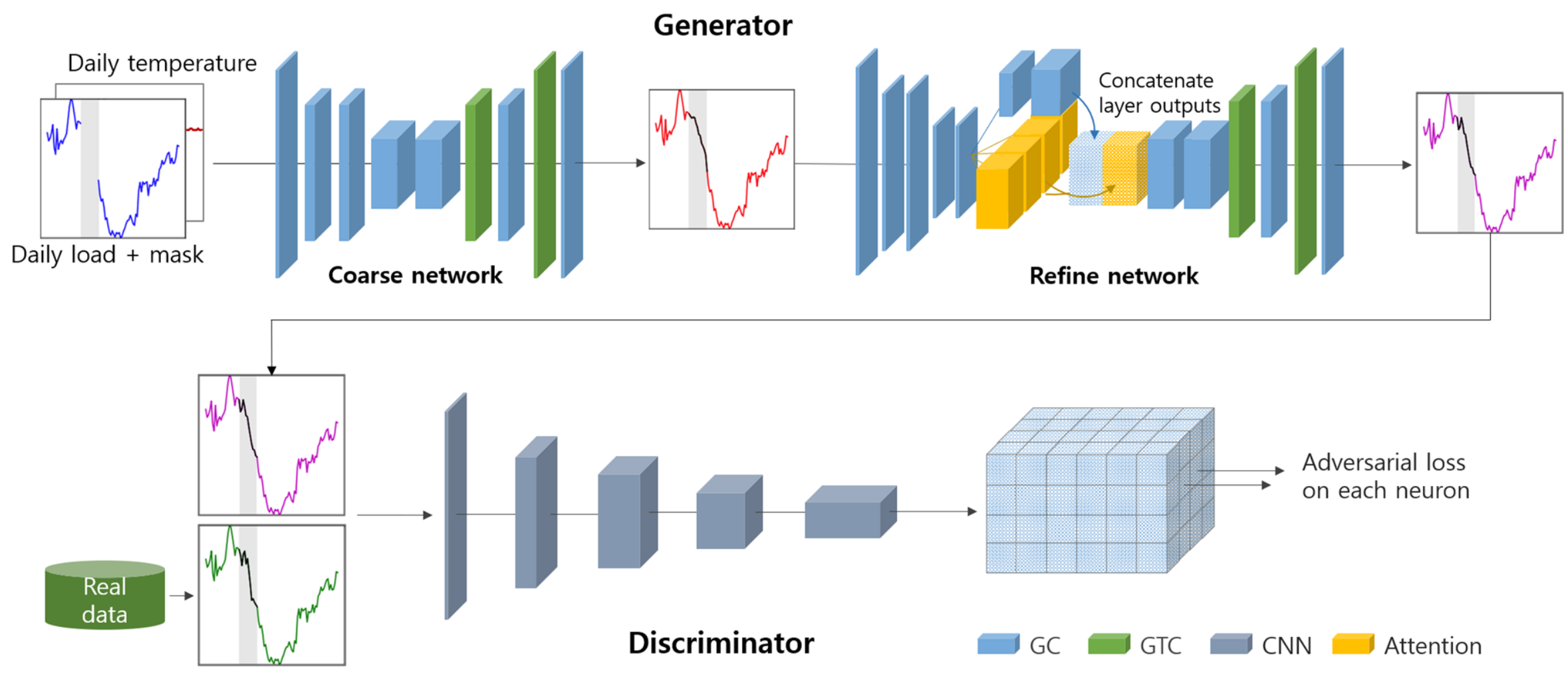

| | | | | | | | | | | | | | | | |
|---|---|---|---|---|---|---|---|---|---|---|---|---|---|---|---|
| **Generator** | **Coarse** | **Layer** | **gc** | **gc** | **gc** | **gc** | **gc** | **gtc** | **gc** | **gtc** | **gc** | | | | |
| | | ks | 5 | 4 | 3 | 4 | 3 | 3 | 3 | 3 | 3 | | | | |
| | | kn | 64 | 128 | 128 | 256 | 256 | 128 | 128 | 64 | 1 | | | | |
| | | st | 1 | 2 | 1 | 2 | 1 | 2 | 1 | 2 | 1 | | | | |
| | **Refine** | **Layer** | **gc** | **gc** | **gc** | **gc** | **gc** | **gc** | **gcn** | **attn*4** | **gcn*2** | **gtcn** | **gcn** | **gtcn** | **gcn** |
| | | ks | 5 | 4 | 3 | 4 | 3 | 3 | 3 | 3 | 3 | 3 | 3 | 3 | 3 |
| | | kn | 64 | 64 | 64 | 64 | 64 | 128 | 256 | 64 | 1 | 2 | 128 | 2 | 128 |
| | | st | 1 | 2 | 1 | 2 | 1 | 1 | 1 | 1 | 256 | 128 | 1 | 128 | 1 |
| **Discriminator** | | **Layer** | **cnn** | **cnn** | **cnn** | **cnn** | **cnn** | | | | | | | | |
| | | ks | 4 | 4 | 4 | 4 | 4 | | | | | | | | |
| | | kn | 16 | 32 | 64 | 128 | 256 | | | | | | | | |
| | | st | 2 | 2 | 2 | 2 | 2 | | | | | | | | |

Fig. 4. The proposed Load-PIN framework. “GC” refers to a gated convolution block, “GTC” refers to a gated transpose convolution block, “CNN” refers to a convolutional block, and “Attention” refers to a self-attention block. “ks” means kernel size, “kn” means number of kernels, and “st” means stride.

The 1st-stage coarse network is trained purely based on the point-to-point content loss function, $L_{coarse}$.

$$L_{coarse} = \frac{1}{H}\left\|G_{\theta_{G1}}(\mathbf{z}) - \mathbf{P}\right\|_2^2 \tag{8}$$

where $\theta_{G1}$ is the parameter of the coarse generator network, $\mathbf{P}$ is the ground truth load profile, and $H$ is the dimension of $\mathbf{P}$.

Note that instead of computing the point-to-point content loss ($L_{coarse}$) for the entire 24-hour period, we calculate only the content loss for the masked segment plus a few points before and after the masked segment, the length of which is $H$. For example, in this paper, $H$ is set as 5 hours to cover the masked segment with a margin of 0.5 hour. This ensures that the coarse network focuses on the masked period to enhance accuracy and achieve smoother transition between non-event periods and the event period (i.e., no spikes during the transitions).

### *3) Generator Network: 2nd-stage Fine-tuning Network*

In the second stage, we make two modifications to the conventional GAN framework: adding *Multi-head self-attention* and *considering the tradeoff between content loss, feature matching loss and adversarial loss in the loss function* to polish the first-stage results and recover realistic details (high-frequency components).

Self-attention, also known as intra-attention, is a technique for focusing attention on various points in a single sequence when creating a representation of the sequence. An attention function can be described as mapping a query and a set of key-value pairs to an output, where the query, keys, values, and output are all vectors. The output is computed as a weighted sum of the value as

$$Attention(\mathbf{Q},\mathbf{K},\mathbf{V}) = softmax(\frac{\mathbf{Q}\mathbf{K}^{\mathrm{T}}}{\alpha})\mathbf{V} \tag{9}$$

where $\mathbf{Q}$, $\mathbf{K}$, and $\mathbf{V}$ include learnable parameters representing query-key-value pairs, respectively, $\alpha$ is a scaling factor. Instead of performing a single attention function, [26] founds it beneficial to linearly project the queries, keys, and values multiple times with different learned linear projections in parallel, which is multi-head attention formulated as

$$MultiHead(\mathbf{Q},\mathbf{K},\mathbf{V}) = concate(head_1, head_2, ..., head_n)\mathbf{W}^{\mathbf{O}} \tag{10}$$
$$head_i = Attention(\mathbf{Q}\mathbf{W}_i^{\mathbf{Q}}, \mathbf{K}\mathbf{W}_i^{\mathbf{K}}, \mathbf{V}\mathbf{W}_i^{\mathbf{V}})$$

where $\mathbf{W}_i^{\mathbf{Q}}$, $\mathbf{W}_i^{\mathbf{K}}$, $\mathbf{W}_i^{\mathbf{V}}$ and $\mathbf{W}^{\mathbf{O}}$ are the learnable parameter matrices of the projection. This allows the temporal dependencies between the elements in the input sequences to be modeled without considering their distances [28].

The loss function of the fine-tuning network ($L_{refine}$) includes 3 terms: the content loss, the adversarial loss ($L_{adv}$) and the feature-matching loss ($L_{feat}$), as shown in (11)-(13). $\lambda_1$ and $\lambda_2$ are the weights. Same as in $L_{coarse}$, the content loss minimizes point-to-point errors. $L_{adv}$ improves the realisticness of the estimation results by maximizing the scores of the discriminator, where $M$ is the dimension of the discriminator output (shown as the blue 3D matrix in Fig. 4). $L_{feat}$ is defined in (13) as the distance between high-level feature maps extracted from the hidden layers of the discriminator network, where $\varphi_j(\cdot)$ represents the output of the $j^{th}$ intermediate convolution layer of the discriminator network. $J$ is the number of intermediate layers in the discriminator network. Because high-level features of real load profiles are embedded in the hidden layer outputs, $L_{feat}$ guides the fine-tuning network to generate more realistic results by matching those high-level features extracted from real profiles. Similar to $L_{coarse}$, all 3 loss terms in $L_{refine}$ are calculated using the 5-hour segment instead of the 24-hour load profile.

$$L_{refine} = \frac{1}{H}\left\|G_{\theta_{G2}}(\mathbf{z}) - \mathbf{P}\right\|_2^2 + \lambda_1 \cdot L_{adv} + \lambda_2 \cdot L_{feat} \quad (11)$$

$$L_{adv} = -\frac{1}{M} D(G_{\theta_{G2}}(\mathbf{z})) \quad (12)$$

$$L_{feat} = \sum_{j=1}^{J}\left\|\varphi_j\left(G_{\theta_{G2}}(\mathbf{z})\right) - \varphi_j(\mathbf{P})\right\|_2^2 \quad (13)$$

*4) Discriminator Network*

The discriminator network is trained to solve the maximization problem defined by (5). In practice, an event can happen at any time of the day with varying lengths. To help the discriminator focus on the event duration, we also send the corresponding Boolean mask together with the load profile. The discriminator contains five convolutional layers with an increasing number of kernels. This allows us to compress the input profiles into high-level feature matrix, in which each element can cover the entire input load profile. Finally, the adversarial loss is applied to each neural to identify the fake and real inputs.

Inspired by [27], we adopt spectral normalization and hinge loss to stabilize the training process of the discriminator by minimizing the loss function $L_D$ calculated as

$$L_D = \frac{1}{M} ReLu(1 - D_{\theta_D}(\mathbf{P})) + \frac{1}{M} ReLU(1 + D_{\theta_D}(G(\mathbf{z}))) \quad (14)$$

where $\theta_D$ is the parameters of the discriminator networks.

*5) Model Performance Evaluation*

Three performance metrics are calculated: normalized Root Mean Squared Error (*nRMSE*), Energy Error (*EE*), and $bias$.

$$nRMSE = \frac{1}{N_{test}} \sum_{i=1}^{N_{test}} \sqrt{T_{event}^{i}\left\|\mathbf{Y}_{event}^{i} - \widehat{\mathbf{Y}}_{event}^{i}\right\|_2^2} \Big/ \left\|\mathbf{Y}_{event}^{i}\right\|_1 \quad (15)$$

$$EE = \frac{1}{N_{test}} \sum_{i=1}^{N_{test}} \left|\sum_{t=1}^{T_{event}^{i}} (y_t^i - \widehat{y}_t^i)\right| \Big/ \left\|\mathbf{Y}_{event}^{i}\right\|_1 \quad (16)$$

$$bias = \frac{1}{N_{test}} \sum_{i=1}^{N_{test}} \left(\frac{1}{T_{event}^{i}} \sum_{t=1}^{T_{event}^{i}} \frac{y_t^i - \widehat{y}_t^i}{y_t^i}\right) \times 100\% \quad (17)$$

where *nRMSE* evaluates the expected point-to-point error; *EE* evaluates the expected accumulated energy error; $bias$ reflects whether the model has a consistent difference between the actual and the missing data.

*6) CVR Efficacy Estimation*

To evaluate the performance of a CVR program containing $N_{CVR}$ CVR events, the trained model is used to estimate the CVR baseline for each CVR sample so the expected load reduction when executing CVR can be calculated.

The raw average, normalized load reduction of the $i^{th}$ CVR event is

$$CVR_{raw}^{i} = \frac{1}{T_{event}^{i}} \sum_{t=1}^{T_{event}^{i}} \frac{y_t^i - \widehat{y}_t^i}{y_t^i} \times 100\% \quad (18)$$

The net average, normalized load reduction of the $i^{th}$ CVR event considering the forecasting bias is

$$CVR_{net}^{i} = CVR_{raw}^{i} - bias \quad (19)$$

When calculating the forecasting bias in CVR baseline estimation, only the normal samples in the same season and having missing data during the CVR event periods are selected, i.e., the bias reflects the consistent difference between the actual and the missing data in CVR periods only.

If $CVR_{net}$ is negative, load is reduced during the CVR period. If $CVR_{net}$ is positive, load is increased during the CVR periods, making the feeder unfit for CVR.

The average CVR factor for all CVR events is

$$CVR_f = \frac{1}{N_{CVR}} \sum_{i=1}^{N_{CVR}} CVR_{net}^{i} \Big/ \Delta V^{i} \quad (20)$$

where $\Delta V^{i}$ is the voltage reduction ratio during the $i^{th}$ CVR event. Note that the CVR factor can be used by utilities to identify suitable feeders for CVR.

## III. Case Study

In this paper, as shown in Table I, three test cases are set up to demonstrate the efficacy of the proposed method: Base case for performance benchmarking, fixed-duration CVR case, and variable-duration CVR case.

TABLE I
Test Case Descriptions

| Case Description | | Data Source | Data Resolution | Data Length | Data Size | CVR Events |
|---|---|---|---|---|---|---|
| 1 | **Base** | PECAN Street | 1-minute | 1-year | 318 residential users | None |
| 2 | **Fixed duration** | Utility A | 15-minute | 2-year | 3 feeders | 24 |
| 3 | **Variable duration** | Utility B | 5-minue | 1-year | 1 substation | 32 |

To benchmark Load-PIN, we compare its performance with five other deep learning models: Multi-layer Perceptron (MLP) [29], Long-short Term Memory (LSTM) [30], Temporal Convolutional Net (TCN) [31], Bi-LSTM (Bidirectional LSTM) [32], and SAE (Stacked Auto-Encoder) [33].

Using only the pre-event data (i.e., $\mathbf{X}_{pre}$ and $\mathbf{Y}_{pre}$) as inputs, MLP, TCN, and LSTM formulate missing data restoration and CVR baseline identification as a load forecasting problem. Bi-LSTM and SAE are bi-direction models. Thus, the inputs for the Bi-LSTM, SAE and Load-PIN models are the same. Model hyper-parameters are selected based on the trial-and-error method.

In each case, all models are trained (70%), validated (15%) and tested (15%) on non-CVR samples. For the two CVR cases, the trained model is applied to the CVR samples for CVR baseline identification. The baselines are then used for computing CVR effects, which will be used to assess the efficacy of the CVR program.

### A. *Performance Evaluation for Missing Data Restoration*

As shown in Table I, in the base case, PECAN Street data [34] is used for performance benchmarking on missing data restoration. To test the impact of data resolution on estimation accuracy, 1-minute data are down-sampled to 5-min, 15-min, 30-min and 1-h. To benchmark the model performance at different load aggregation levels, we test on aggregated load profiles of 10 users, 50 users, 100 users, 200 users and 300 users. After combining 5 data resolutions and 5 load levels, we obtain 25 test scenarios in total. The mask length is set to be 3-hour.

From the results, the following observations are made:

- As shown in Table II, the three bidirectional models (i.e., Bi-LSTM, SAE, and Load-PIN) outperform the one-directional forecasting models in most cases. As shown in Fig. 6, the bidirectional models (solid lines) has lower RMSEs than the forecasting models (dashed lines), especially in the second half of the 3-hour estimation window. The results clearly demonstrate that, by adding post-event data into the input, the accuracy will improve significantly. The step-by-step RMSE distributions shown in Fig. 7 further confirm the above observation.
- As shown in Table II, when data granularity is 5-min and 15-min, Load-PIN outperforms all other models and shows 15-30% improvement compared with the second-best model. This shows that Load-PIN can extract information hidden inside the high-resolution data for forecasting the missing data segments. However, if the data resolution is too low, the Load-PIN does not show significant performance improvements. This is because in those cases, forecasting average values outweigh uncovering load shape details.
- As shown in Fig. 8, when the load aggregation level increases from 10 to 300 users, both $nRMSE$ and $EE$ decrease. As shown in Fig. 8(a), to achieve a smaller $nRMSE$, the best data granularity (highlighted by yellow rectangles) for a group with less than 100 users is 1-h and for 300 users is 15-min. This is because when load aggregation level increases, the load profile becomes smoother. Thus, it is possible to use higher data granularity for restoring more detailed load shapes, which, in turn, improves the point-to-point accuracy. According to Fig. 8(b), to achieve a smaller $EE$, the best data granularity (highlighted by yellow rectangles) is 15-minute.
- As shown in Fig. 9, Load-PIN performance is consistent across different hours of the day and in different seasons. Note that the error is substantially smaller for summer peak load periods (i.e. between 15:00-18:00 in summer months), which is highly desirable for CVR baseline identification.

To summarize, benchmark tests show that Load-PIN outperforms all other algorithms in missing data restoration. Higher resolution data is preferred when there are more users in the group. In most cases, 5 or 15 minutes data set are sufficient for restoring a missing data segment with a duration of 4-hour or less.

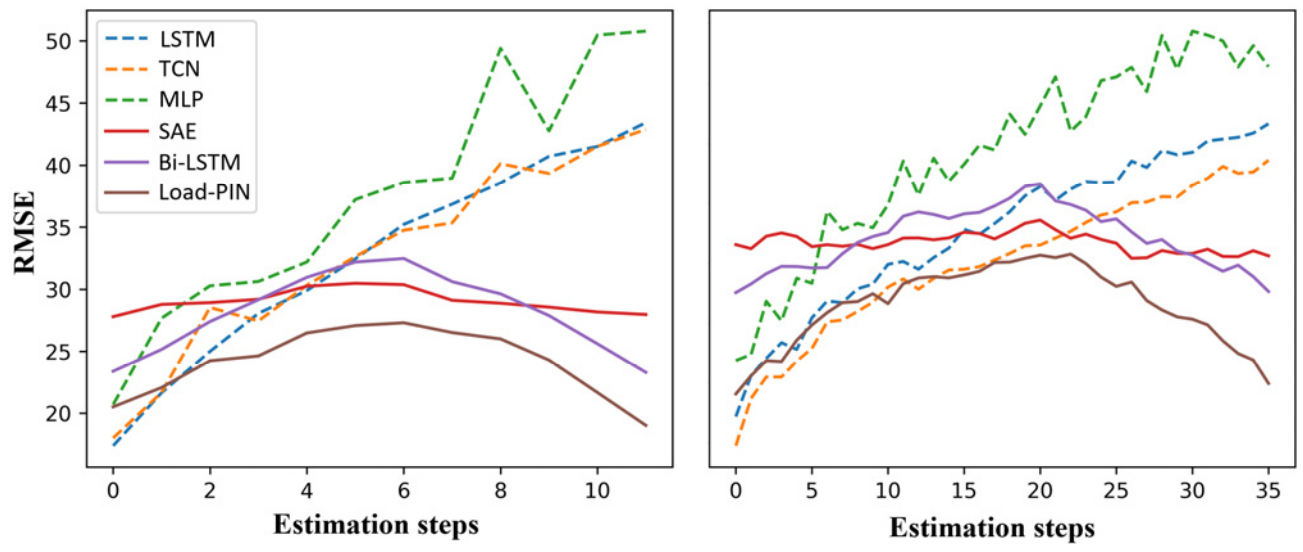

(a) (b)

Fig. 6. Load-PIN performance evaluation: averaged step-by-step RMSE of the 6 models during 3-hour estimation interval. Test on the aggregated load of 300 users. (a) 15-min data granularity, (b) 5-min data granularity.

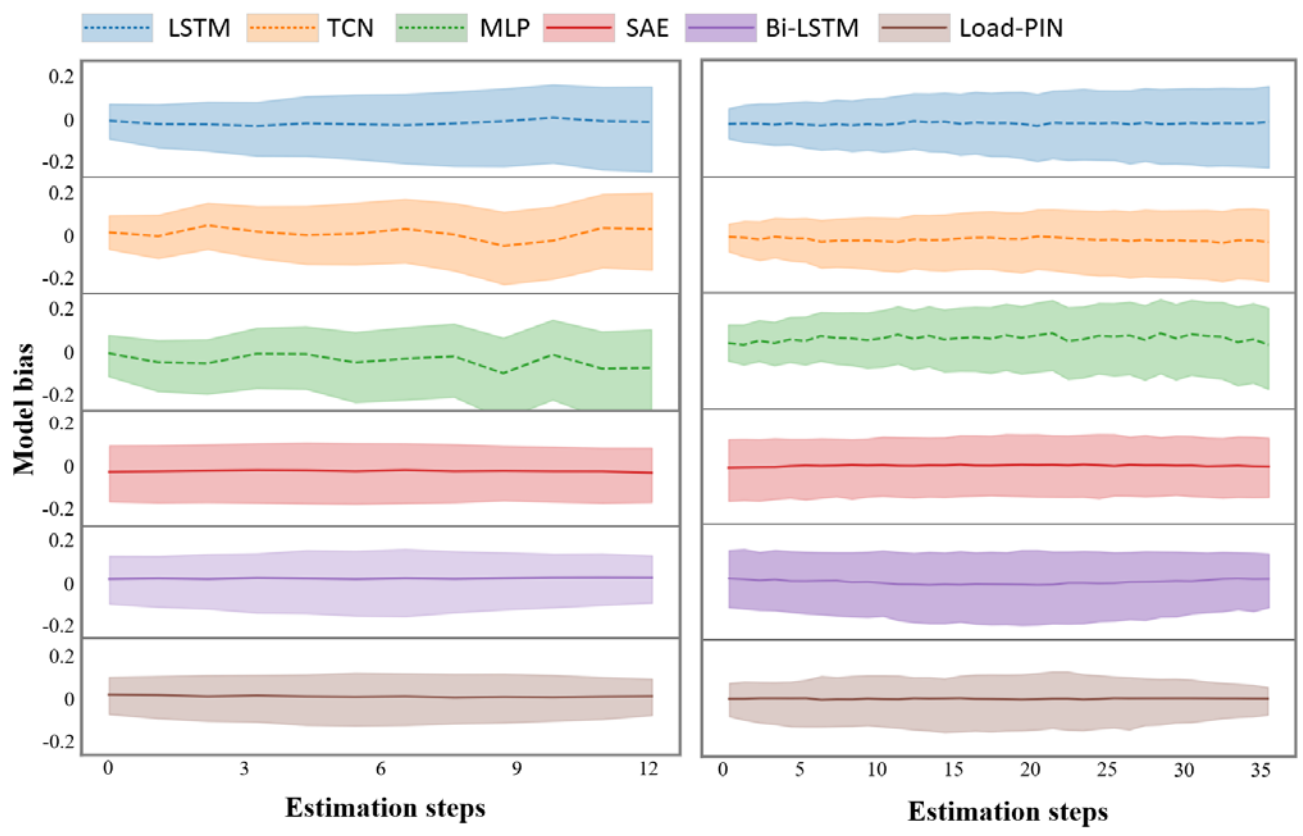

(a) (b)

Fig. 7. Averaged step-by-step RMSE and the 90% confidence intervals of the 6 models during 3-hour estimation period. Test on the aggregated load of 300 users. (a) 15-min data granularity, (b) 5-min data granularity.

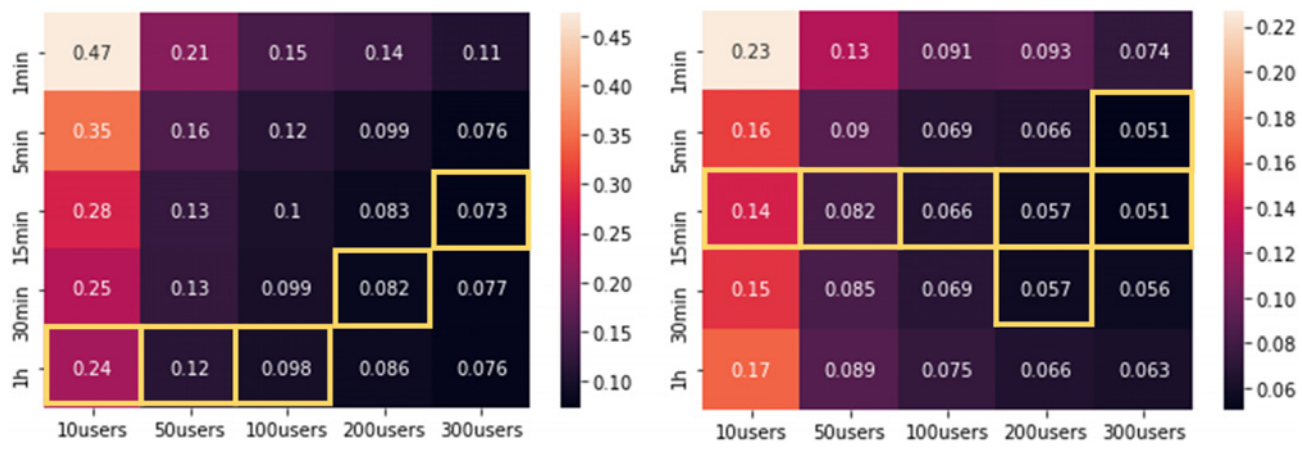

(a) (b)

Fig. 8. Impact of data resolution on missing data restoration. (a) nRMSE, (b) EE. Note the value is an averaged of all 6 deep learning models. The yellow rectangles highlight the smallest error under each aggregation level.

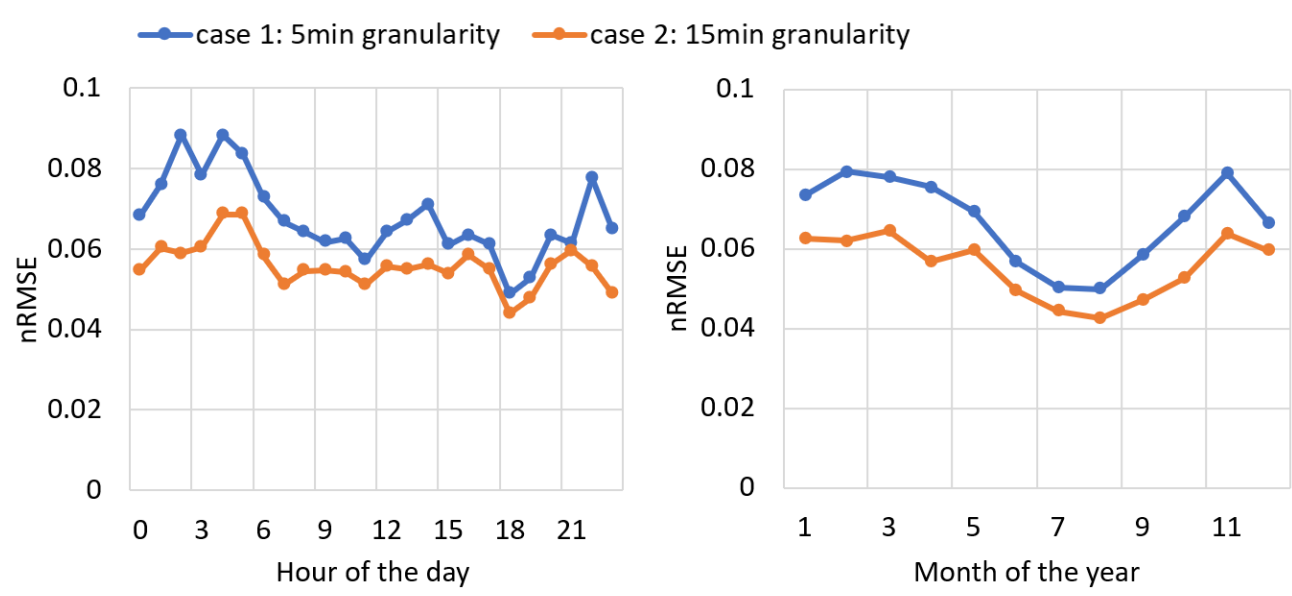

(a) (b)

Fig. 9. Performance of Load-PIN on (a) different hours of the day, (b) months of the year.

TABLE II
Model performances on the Pecan Street Test Case (Mask length is set to be 3 hours)

| data granularity | aggregation level | nRMSE | | | | | | | EE | | | | | | |
|---|---|---|---|---|---|---|---|---|---|---|---|---|---|---|---|
| | | **LSTM** | **TCN** | **MLP** | **SAE** | **Bi-LSTM** | **Load-PIN** | **Improvement** | **LSTM** | **TCN** | **MLP** | **SAE** | **LSTM** | **Load-PIN** | **improvement** |
| **1-min** | 10users | 0.46 | 0.43 | 0.39 | 0.35 | 0.57 | 0.35 | -0.85% | 0.22 | 0.21 | 0.16 | 0.15 | 0.26 | 0.15 | -1.99% |
| | 50users | 0.25 | 0.21 | 0.19 | 0.15 | 0.22 | 0.12 | 18.90% | 0.17 | 0.13 | 0.11 | 0.08 | 0.13 | 0.06 | 23.35% |
| | 100users | 0.19 | 0.17 | 0.15 | 0.11 | 0.16 | 0.10 | 7.79% | 0.13 | 0.10 | 0.09 | 0.06 | 0.10 | 0.06 | 4.48% |
| | 200users | 0.16 | 0.15 | 0.12 | 0.08 | 0.12 | 0.07 | 15.74% | 0.12 | 0.10 | 0.06 | 0.05 | 0.09 | 0.05 | 4.14% |
| | 300users | 0.15 | 0.13 | 0.10 | 0.07 | 0.11 | 0.09 | -25.30% | 0.12 | 0.09 | 0.06 | 0.05 | 0.08 | 0.05 | -12.78% |
| | **mean** | 0.24 | 0.22 | 0.19 | 0.15 | 0.24 | 0.15 | 3.26% | 0.15 | 0.13 | 0.10 | 0.08 | 0.13 | 0.08 | 3.44% |
| **5-min** | 10users | 0.39 | 0.34 | 0.35 | 0.32 | 0.38 | 0.19 | 39.76% | 0.20 | 0.13 | 0.15 | 0.15 | 0.17 | 0.08 | 38.46% |
| | 50users | 0.20 | 0.15 | 0.15 | 0.14 | 0.15 | 0.10 | 26.63% | 0.14 | 0.08 | 0.08 | 0.08 | 0.08 | 0.07 | 9.76% |
| | 100users | 0.15 | 0.11 | 0.12 | 0.10 | 0.11 | 0.11 | -7.36% | 0.10 | 0.06 | 0.07 | 0.06 | 0.06 | 0.05 | 11.79% |
| | 200users | 0.15 | 0.09 | 0.12 | 0.08 | 0.08 | 0.07 | 17.36% | 0.11 | 0.05 | 0.09 | 0.05 | 0.05 | 0.05 | 3.40% |
| | 300users | 0.08 | 0.08 | 0.10 | 0.07 | 0.07 | 0.06 | 13.73% | 0.05 | 0.05 | 0.07 | 0.05 | 0.05 | 0.04 | 24.77% |
| | **mean** | 0.19 | 0.15 | 0.17 | 0.14 | 0.16 | 0.11 | 18.02% | 0.12 | 0.07 | 0.09 | 0.08 | 0.08 | 0.06 | 17.64% |
| **15-min** | 10users | 0.30 | 0.30 | 0.30 | 0.27 | 0.28 | 0.21 | 21.20% | 0.15 | 0.15 | 0.15 | 0.14 | 0.15 | 0.09 | 37.89% |
| | 50users | 0.14 | 0.14 | 0.15 | 0.12 | 0.12 | 0.09 | 24.88% | 0.09 | 0.10 | 0.09 | 0.08 | 0.07 | 0.05 | 28.46% |
| | 100users | 0.11 | 0.11 | 0.12 | 0.09 | 0.09 | 0.08 | 8.68% | 0.07 | 0.07 | 0.09 | 0.06 | 0.06 | 0.05 | 17.67% |
| | 200users | 0.10 | 0.10 | 0.09 | 0.07 | 0.07 | 0.05 | 31.02% | 0.07 | 0.07 | 0.06 | 0.05 | 0.05 | 0.04 | 20.47% |
| | 300users | 0.08 | 0.08 | 0.09 | 0.06 | 0.06 | 0.05 | 15.70% | 0.06 | 0.05 | 0.07 | 0.05 | 0.05 | 0.03 | 35.10% |
| | **mean** | 0.15 | 0.15 | 0.15 | 0.12 | 0.13 | 0.10 | 20.30% | 0.09 | 0.09 | 0.09 | 0.08 | 0.07 | 0.05 | 27.92% |
| **30-min** | 10users | 0.27 | 0.27 | 0.28 | 0.23 | 0.23 | 0.22 | 4.74% | 0.15 | 0.15 | 0.19 | 0.15 | 0.14 | 0.11 | 23.15% |
| | 50users | 0.15 | 0.14 | 0.15 | 0.11 | 0.11 | 0.09 | 15.70% | 0.12 | 0.09 | 0.09 | 0.08 | 0.07 | 0.07 | 2.81% |
| | 100users | 0.11 | 0.12 | 0.12 | 0.08 | 0.08 | 0.08 | 5.02% | 0.08 | 0.09 | 0.08 | 0.06 | 0.06 | 0.05 | 8.16% |
| | 200users | 0.10 | 0.10 | 0.10 | 0.07 | 0.07 | 0.06 | 11.97% | 0.07 | 0.06 | 0.07 | 0.05 | 0.05 | 0.04 | 14.69% |
| | 300users | 0.09 | 0.10 | 0.10 | 0.06 | 0.06 | 0.05 | 16.35% | 0.06 | 0.06 | 0.07 | 0.05 | 0.05 | 0.04 | 12.78% |
| | **mean** | 0.14 | 0.15 | 0.15 | 0.11 | 0.11 | 0.10 | 10.75% | 0.10 | 0.09 | 0.10 | 0.08 | 0.07 | 0.06 | 12.32% |
| **1-h** | 10users | 0.29 | 0.28 | 0.27 | 0.19 | 0.19 | 0.19 | -0.84% | 0.19 | 0.23 | 0.18 | 0.15 | 0.15 | 0.12 | 17.53% |
| | 50users | 0.14 | 0.14 | 0.15 | 0.09 | 0.09 | 0.10 | -11.87% | 0.11 | 0.10 | 0.11 | 0.07 | 0.07 | 0.08 | -8.50% |
| | 100users | 0.13 | 0.13 | 0.11 | 0.07 | 0.07 | 0.07 | 1.72% | 0.10 | 0.09 | 0.08 | 0.06 | 0.06 | 0.05 | 14.47% |
| | 200users | 0.12 | 0.11 | 0.11 | 0.06 | 0.06 | 0.06 | -4.37% | 0.10 | 0.07 | 0.08 | 0.05 | 0.05 | 0.05 | -3.68% |
| | 300users | 0.11 | 0.09 | 0.09 | 0.05 | 0.06 | 0.04 | 21.79% | 0.09 | 0.06 | 0.07 | 0.04 | 0.05 | 0.04 | 6.43% |
| | **mean** | 0.16 | 0.15 | 0.15 | 0.09 | 0.09 | 0.09 | 1.28% | 0.12 | 0.11 | 0.10 | 0.07 | 0.08 | 0.07 | 5.25% |

### *B. Performance Evaluation on CVR Baseline Estimation*

In this section, we apply the proposed Load-PIN model to estimate the CVR baseline for actual feeders in North Carolina, USA. For each feeder, we first train and test the Load-PIN model using data collected in non-CVR days (same as in Section III.A). Next, the trained model is used for baseline estimation in CVR days. The estimated baselines are used for CVR efficacy assessment.

#### *1) Fixed-duration CVR Case*

As shown in Table 1, the fixed CVR duration case is conducted using data collected from three residential distribution feeders in utility A, namely BR, DF and SL. All 24 labeled CVR events are in summer months with a fixed duration of 3 hours between 14:00 and 19:00 p.m. The CVR voltage reduction is 4%. The model performance is evaluated on non-CVR days. Note that forecasting bias is removed using methods introduced in Section II.C.6 when calculating CVR load reduction.

TABLE III
MODEL PERFORMANCES ON THE 3 TEST FEEDERS (IN PERCENTAGE)

| (%) | Feeder | TCN | LSTM | MLP | SAE | BLSTM | Load-PIN |
|---|---|---|---|---|---|---|---|
| **nRMSE** | **BR** | 3.12 | 3.13 | 3.8 | 4.05 | 3.67 | 2.50 |
| | **DF** | 3.87 | 3.78 | 5.56 | 6.35 | 5.8 | 2.86 |
| | **SL** | 3.66 | 3.72 | 3.91 | 4.03 | 3.67 | 3.02 |
| **EE** | **BR** | 2.02 | 2.03 | 2.63 | 2.91 | 2.54 | 1.13 |
| | **DF** | 2.86 | 2.75 | 4.03 | 4.91 | 3.78 | 1.98 |
| | **SL** | 2.54 | 2.65 | 2.77 | 2.47 | 2.18 | 1.40 |
| **Absolute Bias** | **BR** | 0.33 | 0.37 | 1.52 | 1.79 | 0.39 | 0.03 |
| | **DF** | 1.63 | 0.18 | 2.34 | 3.32 | 1.13 | 0.22 |
| | **SL** | 1.03 | 1.21 | 1.1 | 1.31 | 0.11 | 0.05 |

As shown in Table III, Load-PIN outperforms the other methods (lowest $nRMSE$ and $EE$) by a large margin. From the results, we made the following observations

- As shown in Fig. 10, the results obtained by Load-PIN (the solid brown line) show similar trends as the average of all six models (the solid pink line). This shows that Load-PIN captures the trending information well.
- In Fig. 11, we randomly plot four CVR baselines for each feeder (out of 24 CVR events). The forecasted baseline shows a smooth transition from the pre-CVR to CVR periods and from the CVR to post-CVR periods.
- As shown in Figs. 10 and 11, for all three feeders, we observe clear load reduction (i.e. the value of CVR effect is negative in Fig. 10) in the first 1.5-hour. However, for feeders BR and DF, a pay-back period is observed in the second 1.5-hour. In fact, the amount of load reduction starts to diminish after the first hour. In some cases, feeder loads start to rise after the initial drop until exceeding the baseline.
- This CVR diminishing and subsequent pay-back effect is very likely caused by the increasing penetration of thermostatically controlled appliances (e.g., refrigerators, ovens, air conditioners) and LED lighting loads, which are rapidly replacing the incandescent lighting loads. Note that when voltage is low, appliances will turn on longer if a fixed amount of energy is required in each duty cycle.
- For feeder SL, even though the pay-back effect is less visible compared to feeders BR and DF, the CVR effect

also diminishes after one hour.

- Overall, the results show that CVR efficacy will diminish in a prolonged CVR event, which indicates that the utilities may need to execute CVR for a period less than 2-hour.

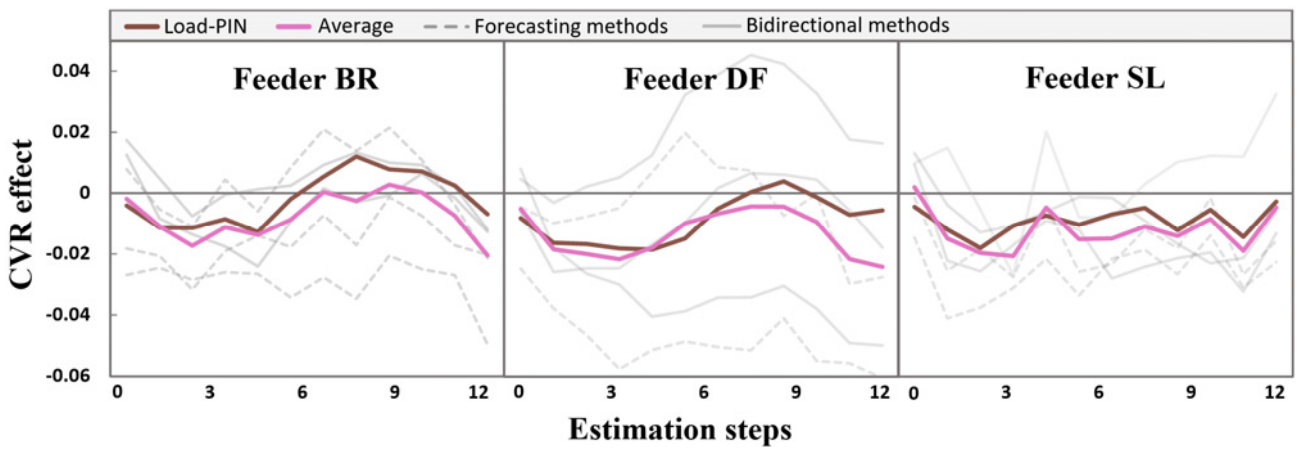

Fig. 10. Averaged step-by-step CVR effects of the 3 test feeders.

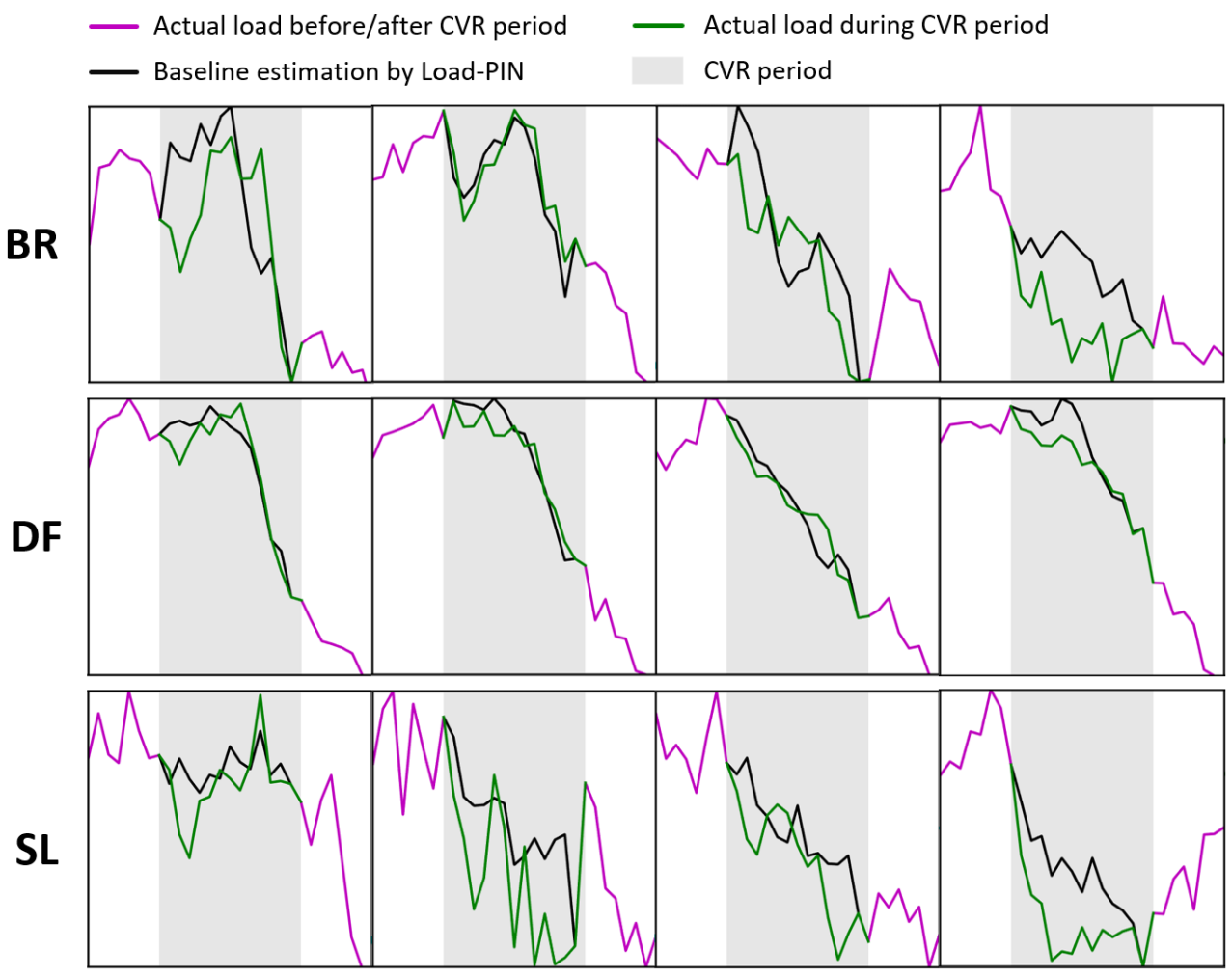

Fig. 11. Examples of the Load-PIN generated CVR baseline.

### 2) *Variable-duration CVR Case*

As shown in Table I, in this case, we use 5-minute data collected by a utility at a distribution substation bus to identify the CVR baseline for CVR events with variable durations. There are 32 CVR events in summer with durations ranging from 75 minutes up to 190 minutes. Figure 12 shows the daily voltage profiles at the feeder head of 3 identified CVR days as an example.

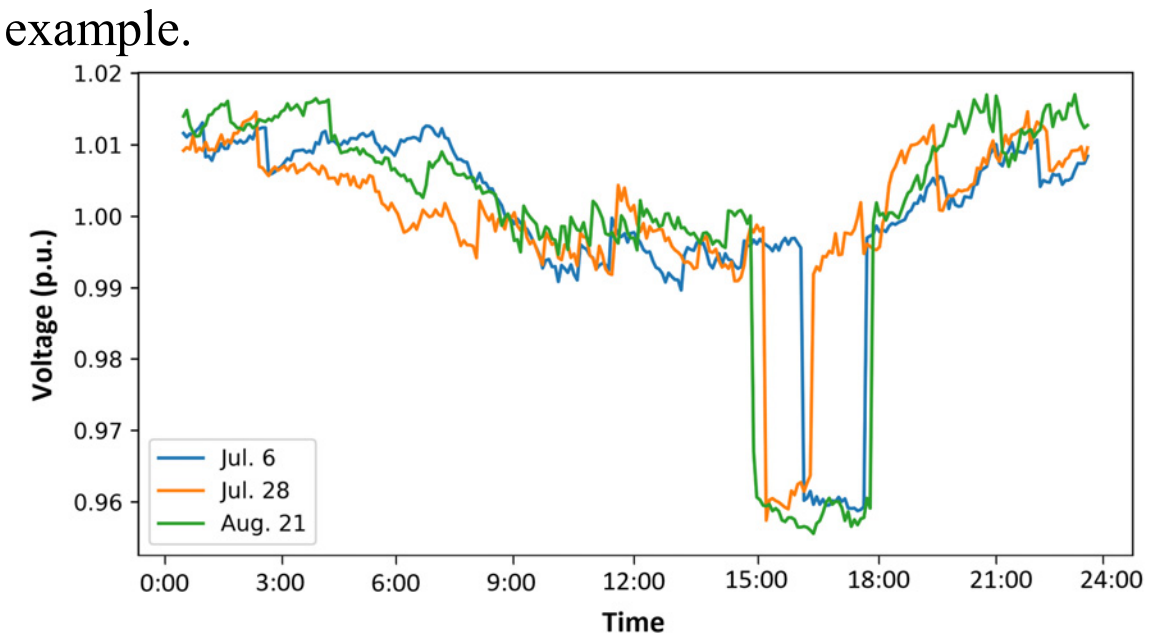

Fig. 12. Examples of daily voltage profiles of 3 CVR event days measured at the feeder head.

To make the bidirectional models (including Load-PIN) adapt to the varying-length CVR events, we randomly put a mask to each training sample with varying lengths between 1 - 4 hours. For the 3 forecasting models, we fix the output window to 4 hours so all possible CVR durations are covered in their forecasting range.

As shown in Table IV, Load-PIN outperforms all other models by achieving the smallest *nRMSE* and *EE*. The model bias is 0.43 only slightly higher than that of LSTM. The results demonstrate that Load-PIN can handle varying-length estimation tasks by leveraging the closest bidirectional data around each CVR events.

TABLE IV
MODEL PERFORMANCE ON VARYING-LENGTH FEEDER CASE (IN PERCENTAGE)

| (%) | TCN | LSTM | MLP | SAE | BLSTM | Load-PIN |
|---|---|---|---|---|---|---|
| **nRMSE** | 3.23 | 4.13 | 4.93 | 4.31 | 3.18 | **2.15** |
| **Energy** | 1.91 | 2.69 | 3.37 | 3.03 | 1.95 | **1.07** |
| **Absolute Bias** | 0.82 | **0.27** | 1.17 | 1.74 | 1.57 | 0.43 |

We implement the 6 trained models to estimate the CVR baseline for all the 32 events, and then calculate the averaged CVR effect, as shown in Fig. 13. Note that due to the uneven durations of the CVR events, results are less stable after 2h because the durations of many CVR events are less than 2 hours. However, we can still see similar trends with the fixed-duration case (using data from utility B): the CVR effect is more significant in the first hour and the initial load drop period will be followed by a pay-back period where load will rise.

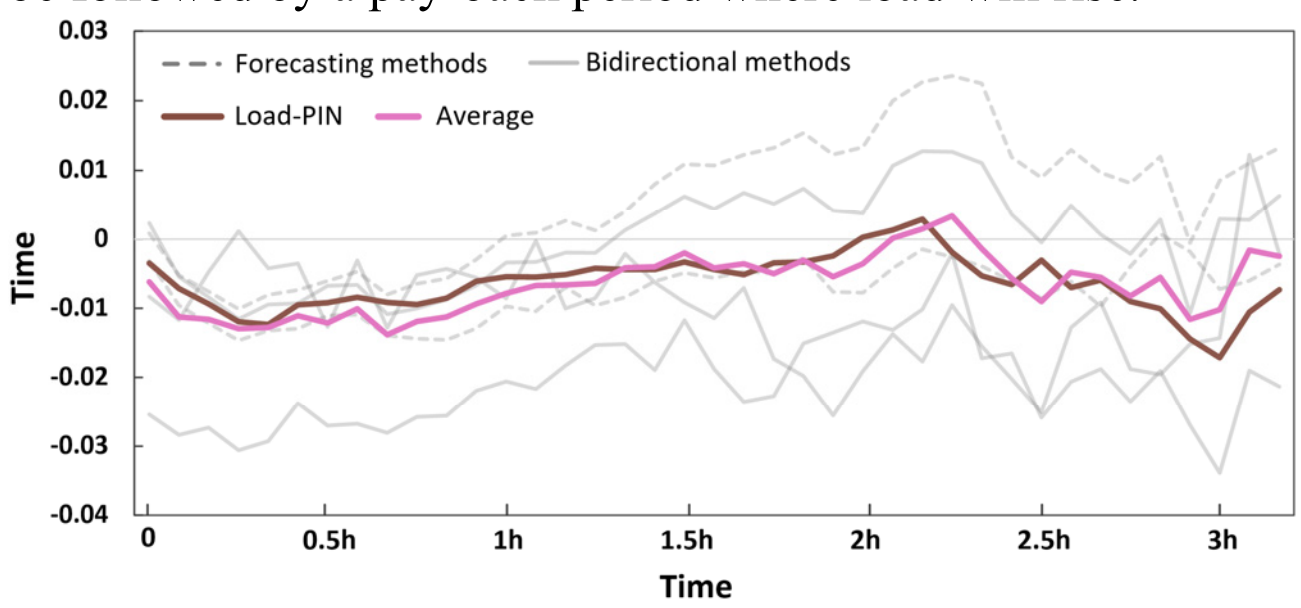

Fig. 13. Averaged step-by-step CVR effects at a substation bus.

## IV. CONCLUSION

In this paper, we propose a novel deep-learning model Load-PIN to solve the missing data restoration and CVR baseload estimation problem. Load-PIN merges Gated Convolution and Multi-head self-attention mechanisms into the GAN based framework to enhance the estimation accuracy. Load-PIN is trained using the dynamic-masking strategy so that it can handle CVR events with varying durations. We first demonstrate that at higher load aggregation levels, higher data resolution can achieve better estimation accuracy. In general, 5-min and 15-min resolutions are sufficient for feeder level studies. Next, we demonstrate that the Load-PIN model can achieve 15-30% accuracy improvement under the suggested data granularity, compared with 5 benchmarking methods. Using the trained Load-PIN model for CVR baseline identification, we computed the CVR effects for CVR programs with fixed- and variable-durations. We show that CVR can achieve load reduction in the first 1 hour. However, after 1.5 hours, the CVR effect starts to diminish and a pay-pack period can be observed. This may cause unexpected load peaks in post-CVR periods.

From the results, we want to make two recommendations. First, the CVR execution duration should be less than 2-hours. Second, feeders with high penetration of thermostatically controlled loads may not be good candidates for prolonged CVR programs.